\documentstyle[12pt]{article}
\newcommand{\vf}{\varphi}
\newcommand{\be}{\begin{equation}}
\newcommand{\ee}{\end{equation}}
\newcommand{\ba}{\begin{eqnarray}}
\newcommand{\ea}{\end{eqnarray}}
\newcommand{\no}{\nonumber\\}
\newcommand{\bi}{\bibitem}
\begin{document}
{\it IFA-FT-434-1998}
\bigskip\bigskip
\begin{center}
{\bf \Large{SUPERSYMMETRIES AND CONSTANTS OF \\
~\\
MOTION IN TAUB-NUT SPINNING SPACE}}
\end{center}
\vskip .5truecm
\centerline{{\bf\large
{Diana Vaman\footnote{Present address: Dept. of Physics, SUNY at Stony 
Brook NY 11794, USA;~\\
~E-mail:dvaman@insti.physics.sunysb.edu}
~~{\it and}~~
{\bf\large Mihai Visinescu\footnote
{E-mail: mvisin@theor1.ifa.ro}}}}}
\vskip5mm
\centerline{Department of Theoretical Physics}
\centerline{Institute of Atomic Physics, P.O.Box MG-6, Magurele,}
\centerline{Bucharest, Romania}                                      
\vskip 1cm
\bigskip 
\nopagebreak \begin{abstract}
\noindent
We review the geodesic motion of pseudo-classical spinning particles in
curved spaces. Investigating the generalized Killing equations for
spinning spaces, we express the constants of motion in terms of 
Killing-Yano tensors. The general
results are applied to the case of the four-dimensional
Euclidean Taub-NUT spinning space. A simple exact solution,
corresponding to trajectories lying on a cone, is given..
\end{abstract}

\vskip .2cm
PACS number(s): ~~~~04.20.Jb,~~02.40.-K
\vskip .2cm
\newpage
\section{Introduction}
The models of relativistic particles with spin have been proposed for a
long time. The first published work concerning the lagrangian
description of the relativistic particle with spin was the paper
by Frenkel which appeared in 1926 [1]. After that the
literature on the particle with spin grew vast [2].

The models involving only conventional coordinates are called the
classical models while the models involving anticommuting
coordinates are generally called pseudo-classical.

In this paper we shall confine ourselves to discuss the
relativistic spin one half particle models involving
anticommuting vectorial degrees of freedom which are usually
called the spinning particles. Spinning particles are in some
sense the classical limit of the Dirac particles. After the first
quantization these new anticommuting variables are mapped into
the Dirac matrices and they disappear from the theory [3,4].

The action of spin one half relativistic particle with spinning
degrees of freedom described by Grassmannian (odd) variables was
first proposed by Berezin and Marinov [5] and soon after that
was discussed and investigated in the papers [6-10].

In spite of the fact that the anticommuting Grassmann variables do
not admit a direct classical interpretation, the lagrangians of these
models turn out to be suitable for the path integral description of
the quantum dynamics. The pseudo-classical equations acquire physical
meaning when averaged over inside the functional integrals [5,11].
In the semi-classical regime, neglecting higher order quantum
correlations, it should be allowed to replace some combinations of
Grassmann spin variables by real numbers. Using these ideas the motion 
of spinning particles in external fields have been studied in Refs.
[5, 12-14].
 
On the other hand, generalizations of Riemannian geometry based on
anticommuting variables have been proved to be of mathematical interest.
Therefore the study of the motion of the spinning particles in curved 
space-time is well motivated.

In the present paper we investigate the motion of pseudo-classical
spinning point particles in curved spaces. The generalized
Killing equations for the configuration space of spinning
particles (spinning space) are analysed and the solutions of the
homogeneous part of these equations are expressed in terms of
Killing-Yano tensors. We mention that the existence of a
Killing-Yano tensor is both a necessary and a sufficient
condition for the existence of a new supersymmetry for the
spinning space [15-17]. 

The general results are applied to the case of the
four-dimensional Euclidean Taub-NUT spinning space. The
motivation to carry out this example is twofold. First of all, in the Taub-NUT
geometry there are known to exist four Killing-Yano tensors [18]. From this 
point of view the spinning Taub-NUT space is an exceedingly interesting space 
to exemplify the effective construction of all conserved quantities in terms of
geometric ones, namely Killing-Yano tensors. On the other hand, the Taub-NUT
geometry is involved in many modern studies in physics. For example
the Kaluza-Klein monopole of Gross and Perry [19] and of Sorkin [20] was 
obtained by embedding the Taub-NUT gravitational instanton into 
five-dimensional Kaluza-Klein theory. Remarkably the same object has 
re-emerged in the study of monopole scattering. In the long distance limit, 
neglecting radiation, the relative motion of slow Bogomolny-Prasad-Sommerfield
monopoles is described by the geodesics of this space [21,22].  
The dynamics of well-separated monopoles is completely soluble and has a Kepler
type symmetry [18,23-25]. The problem of geodesic motion in this metric has
therefore its own interest, independently of monopole scattering.

The plan of this paper is as follows. In Sec. 2 we summarize the 
relevant equations for the motions of spinning points in curved 
spaces. The generalized Killing equations for spinning spaces are investigated
and the constants of motion are derived in terms of the solutions of these 
equations.
In Sec. 3 we analyse the motion of pseudo-classical spinning 
particles  in the Euclidean Taub-NUT space. We examine the generalized
Killing equations for this spinning space and describe the derivation of 
the constants of motion in terms of the Killing-Yano tensors.
In Sect. 4 we solve the equations given in the previous 
Section for the special case of motion on a cone. This case represents an 
extension of the scalar particle motions in the usual Taub-NUT space in 
which the orbits are conic sections [18,23-25]. An explicit exact solution 
is given and, in spite of its simplicity, this solution is far from trivial.
Our comments and concluding remarks are presented in Sec. 5.
\section{Spinning spaces and Killing equations}
Spinning particles, such as Dirac fermions, can be described by 
pseudo-classical mechanics  models involving anticommuting c-numbers 
for the spin degrees of freedom. The configuration space of spinning 
particles (spinning space) is an extension of an ordinary Riemannian 
manifold, parametrized by local 
coordinates {$\{$}$x^\mu${$\}$}, to a graded manifold parametrized by local 
coordinates {$\{$}$x^\mu, \psi^\mu${$\}$}, with the first set of variables 
being Grassmann-even (commuting) and the second set Grassmann-odd 
(anticommuting) [3-17].

The dynamics of spinning point-particles in a curved space-time is
described by the one-dimensional $\sigma$-model with the action:
\be
 S=\int_{a}^{b}d\tau \left(\,{1\over 2}\,g_{\mu \nu}(x)\,\dot{x}^\mu 
\,\dot{x}^\nu\, +\, {i\over 2}\, g_{\mu \nu}(x)\,\psi^\mu \,{D\psi^\nu\over
D\tau} \right). 
\ee
Here and in the following, the overdot denotes an ordinary proper-time 
derivative $d/d\tau$, whilst the covariant derivative of $\psi^\mu$
is defined by
\be
{D\psi^\mu\over D\tau}=\dot{\psi}^\mu+\dot{x}^\lambda \,\Gamma^\mu
_{\lambda\nu}\,\psi^\nu .  
\ee

The trajectories, which make the action stationary under arbitrary 
variations $\delta x^\mu $ and $\delta \psi^\mu$ vanishing at the end 
points, are given by:
\ba
{D^2 x^\mu\over D\tau^2}&=&\ddot x^{\,\mu} + \Gamma^\mu_{\lambda\nu}
\,\dot{x}^\lambda\,\dot{x}^\nu={1\over 2i}\,\psi^\kappa\,\psi^\lambda
\,R^{~~\mu~}_{\kappa\lambda~\nu}\,\dot{x}^\nu,\\
{D\psi^\mu\over D\tau}&=&0 .
\ea

The antisymmetric tensor
\be
S^{\mu\nu}=-i\,\psi^\mu\psi^\nu 
\ee
can formally be regarded as the spin polarization tensor of the particle.
The first equation of motion (3) implies the existence of a spin dependent
gravitational force [14]
\be
{D^2 x^\mu\over D\tau^2}={1\over 2}S^{\kappa\lambda}
R^{~~\mu~}_{\kappa\lambda~\nu}\,\dot{x}^\nu
\ee
which is analogous to the electromagnetic force, with spin replacing the
electric charge as the coupling constant. The second equation of motion (4) 
can be expressed in terms of this tensor (5)
and it asserts that the spin is covariantly constant
\be
{DS^{\mu\nu}\over D\tau}=0 . 
\ee

The interpretation of $S^{\mu\nu}$ as spin tensor is corroborated by studying
electromagnetic interaction of the particle [5,9,13,14]. From such an analysis
it results that the space-like components are proportional to the magnetic
dipole moment of the particle, whilst the time-like components $S^{0i}$
represent the electric dipole moment. The requirement that for free Dirac
particles the electric dipole moment vanishes in the rest frame can be written
as a covariant constraint [3]
\be
g_{\nu\lambda} S^{\mu\nu}\dot{x}^\lambda = 0
\ee
which, in terms of the Grassmann coordinates, it is equivalent to
\be
g_{\mu\nu} \dot{x}^\mu \psi^\nu = 0.
\ee

The concept of Killing vector can be generalized to the case of spinning 
manifolds. For this purpose we consider the world-line hamiltonian given by
\be
H=\frac1 2 g^{\mu\nu}\Pi_\mu \Pi_\nu
\ee
where 
\be
\Pi_\mu = g_{\mu\nu}\dot{x}^\nu
\ee
is the covariant momentum.

For any constant of motion ${\cal J}(x,\Pi,\psi)$, the bracket
with $H$ vanishes
\be
\left\lbrace H,{\cal J} \right\rbrace = 0
\ee
where the Poisson-Dirac brackets for functions of the covariant 
phase space variables $(x,\Pi,\psi)$ is defined by
\be
\left\lbrace F,G\right\rbrace={\cal D}_\mu F\frac{\partial G}
{\partial
\Pi_\mu} - \frac{\partial F}{\partial \Pi_\mu}{\cal D}_\mu G -
{\cal R}_{\mu\nu}\frac{\partial F}{\partial
\Pi_\mu}\frac{\partial G}{\partial \Pi_\nu} +
i(-1)^{a_F}\frac{\partial F}{\partial \psi^\mu}\frac{\partial
G}{\partial \psi_\mu}.
\ee
The notations used are
\be
{\cal D}_\mu F = \partial_\mu F +
\Gamma^\lambda_{\mu\nu}\Pi_\lambda\frac{\partial F}{\partial
\Pi_\nu} - \Gamma^\lambda_{\mu\nu}\psi^\nu \frac{\partial
F}{\partial \psi^\lambda}~~~;~~~
{\cal R}_{\mu\nu} = \frac{i}{2}\psi^\rho\psi^\sigma 
R_{\rho\sigma\mu\nu}
\ee
and $a_F$ is the Grassmann parity of $F$ : $a_F=(0,1)$ 
for $F$=(even,odd).

If we expand ${\cal J}(x,\Pi,\psi)$ in a power series in the
covariant momentum 
\be
{\cal J}=\sum_{n=0}^{\infty}\frac{1}{n!}{\cal J}^{(n)\mu_1
\dots\mu_n}(x,\psi) \Pi_{\mu_1}\dots\Pi_{\mu_n}
\ee
then the bracket $\{ H , {\cal J}\}$ vanishes for
arbitrary $\Pi_\mu$ if and only if the components of ${\cal J}$ 
satisfy the generalized Killing equations [3,15,26] :
\be
{\cal J}^{(n)}_{(\mu_1\dots\mu_n;\mu_{n+1})} + \frac{\partial
{\cal J}^{(n)}_{(\mu_1 \dots\mu_n}}{\partial \psi^\sigma}
\Gamma^\sigma_{\mu_{n+1})\lambda} \psi^\lambda = 
\frac{i}{2}\psi^\rho \psi^\sigma R_{\rho\sigma\nu(\mu_{n+1}}
{{\cal J}^{(n+1)\nu}}_{\mu_1 \dots \mu_n)}
\ee
where the parentheses denote symmetrization with norm one over the 
indices enclosed.

In the scalar case, neglecting the Grassmann
variables $\{ \psi^\mu \}$, all the generalized Killing
equations (16) are homogeneous and decoupled. The first equation ($n=0$)
shows that ${\cal J}^{(0)}$ is a trivial constant, the next one ($n=1$) is
the equation for the Killing vectors and so on. In general, for a given 
$n$, neglecting all spin degrees of freedom, eq.(16) defines a Killing 
tensor of valence $n$
\be
{\cal J}^{(n)}_{(\mu_1\dots\mu_n;\mu_{n+1})}(x) = 0
\ee
and from eq.(15) 
\be
{\cal J} ={\cal J}^{(n)}_{\mu_1
\dots\mu_n}(x) \Pi^{\mu_1}\dots\Pi^{\mu_n}
\ee
is a first integral of the geodesic equation [27].

In the spinning case the symmetries can be divided
into two classes.  First, there are four independent {\it generic}
symmetries  which exist in any theory :

\begin{enumerate}
\item{Proper-time translations generated by the hamiltonian $H$ (10);}
\item{Supersymmetry generated by the supercharge
\be
Q_0=\Pi_\mu\,\psi^\mu;
\ee}
\item{Chiral symmetry generated by the chiral charge
\be
\Gamma_* = \frac{i^{[\frac{d}{2}]}}{d!}\sqrt{g}\epsilon_{\mu_1 
\dots \mu_d} \psi^{\mu_1} \dots \psi^{\mu_d};
\ee}
\item{Dual supersymmetry generated by the dual supercharge
\be
Q^* =i\{ \Gamma_* , Q_0 \}
=\frac{i^{[\frac{d}{2}]}}{(d-1)!}\sqrt{g}\epsilon_{\mu_1 \dots 
\mu_d} \Pi^{\mu_1}\psi^{\mu_2} \dots \psi^{\mu_d}
\ee}
\end{enumerate}
where $d$ is the dimension of space-time.

As a rule we have the freedom to choose the value of the supercharge 
$Q_0$ and any choice gives a consistent model. The condition for the
absence of an intrinsic electric dipole moment of physical fermions
(leptons and quarks) as formulated in eq.(9) implies
\be
Q_0 = 0.
\ee
However, for the time being, we shall not fix the value of the
supercharge, keeping the presentation as general as possible.

The second kind of conserved quantities, called {\it
non-generic}, depend on the explicit form of the metric
$g_{\mu\nu}(x)$. In the recent literature there are exhibited the
constants of motion in the Schwarzschild [28], Taub-NUT 
[17,29-33], Kerr-Newman [15,16] spinning spaces.

In what follows we shall deal with the {\it non-generic}
constants of motion in connection with the Killing eqs.(16)
looking for the general features of the solutions. The spinning particle 
constants of motion  can be seen either as extensions of the constants 
from the scalar case or new ones depending on the Grassmann-valued
spin variables {$\{$}$\psi^\mu${$\}$}.
 
Let us assume that the number of terms in the series (15) is
finite. That means that, for a given $n$, 
~${\cal J}^{(n+1)}_{\mu_1 \dots\mu_{n+1}}$
vanishes and the last non-trivial generalized Killing equation from 
the system (16) is in fact homogeneous :
\be
{\cal J}^{(n)}_{(\mu_1 \dots\mu_n;\mu_{n+1})}+
\frac{\partial {\cal J}^{(n)}_{(\mu_1 \dots\mu_n}}
{\partial \psi^\sigma}
\Gamma^\sigma_{\mu_{n+1})\lambda}\psi^\lambda = 0.
\ee

The line of action to solve the system of coupled differential equations 
(16) is standard. One starts with a ${\cal J}^{(n)}_{\mu_1 \dots
\mu_n}$ solution of the homogeneous eq.(23) which has to be
introduced in the right-hand side (RHS) of 
the generalized Killing equations (16) for ${\cal J}^{(n-1)}_{\mu_1
\dots\mu_{n-1}}$ and the iteration must be carried on to $n=0$.

For the beginning let us note that eq.(23) has solutions which do not
depend on the Grassmann coordinates. These are the Killing tensors of
valence $n$, as it can be seen comparing eq.(17) with eq.(23) in which all
spin degrees of freedom are neglected. However, for the spinning
particles, the generalized Killing equations (16) are not decoupled.
Even if one starts with a Killing
tensor of valence $n$ we get from the remaining Killing equations 
the components 
${\cal J}^{(m)}_{\mu_1\dots\mu_m} (m<n)$ with 
non-trivial spin contributions. 

Therefore the quantity (18) is no more conserved and
the actual constant of motion is
\be
{\cal J}=\sum_{m=0}^{n}\frac{1}{m!}{\cal J}^{(m)}_{\mu_1
\dots\mu_m}(x,\psi) \Pi^{\mu_1}\dots\Pi^{\mu_m}
\ee
in which ${\cal J}^{(m)}_{\mu_1\dots\mu_m}(x,\psi)$ with $m<n$ has a
non-trivial spin dependent expression.

The construction of the conserved quantity (24) in which the
last term ${\cal J}^{(n)}_{\mu_1\dots\mu_n}$ is a Killing tensor
can be done effectively.
We shall illustrate this construction with a few examples.
For $n=0$ eq.(17) is satisfied by a simple, irrelevant
constant. The first non-trivial case is $n=1$. In this case eq.(17) 
is satisfied by a Killing vector  $R_\mu$: 
\be
 R_{(\mu;\nu)} = 0.
\ee

Introducing this Killing vector in the RHS of the generalized
Killing eq.(16) for $n=0$ one obtains for the
${\cal J}^{(0)}$ the expression [17]:
\be
{\cal J}^{(0)} = \frac i2 R_{[\mu;\nu]} \psi^\mu \psi^\nu 
\ee
where the square bracket denotes antisymmetrization with norm one.
Consequently, starting with a Killing vector $R_\mu$, we get in the
spinning case the conserved quantity (24) in the form
\be
{\cal J} = \frac i2 R_{[\mu;\nu]} \psi^\mu \psi^\nu + R_\mu \Pi^\mu.
\ee

A more involved example is given by a Killing tensor 
${\cal J}^{(2)}_{\mu\nu} = K_{\mu\nu}$  
satisfying eq.(17) for $n=2$:
\be
K_{(\mu\nu;\lambda)} = 0.
\ee

This solution must be introduced in the RHS of the generalized Killing
equation (16) for ${\cal J}^{(1)}_\mu$ and then we have to evaluate the new 
${\cal J}^{(0)}$.
Unfortunately it is not possible to find closed, analytic
expressions for ${\cal J}^{(1)}_\mu$ and ${\cal J}^{(0)}$ involving the 
spin variables using directly the components of the Killing tensor 
$K_{\mu\nu}$ . But assuming that the Killing tensor
$K_{\mu\nu}$ can be written as a symmetrized product of two
Killing-Yano tensors, the construction of the conserved quantity
(24) is feasible. 

We remind that a tensor  $f_{\mu_1 \dots\mu_r}$ 
is called a Killing-Yano tensor of valence $r$  [27,34] if it 
is totally antisymmetric and it satisfies the equation 
\be
f_{\mu_1 \dots\mu_{r-1}(\mu_{r};\lambda)} = 0.
\ee

It is known that the Killing-Yano tensors play a key role in the Dirac
theory on a curved space-time [35]. The study of the generalized
Killing equations strengthens the connection of the Killing-Yano 
tensors with the supersymmetric classical and quantum mechanics on 
curved manifolds.

For the generality, let us assume that the Killing tensor $ K_{\mu\nu}$
can be written as a symmetrized product of two different Killing-Yano 
tensors
\be
 K^{\mu\nu}_{ij} = {1\over 2}(f^\mu_{i~\lambda} f^{\nu\lambda}_j +
f^\nu_{i~\lambda} f^{\mu\lambda}_j)
\ee
where $f^{\mu\nu}_{i}$ is a Killing Yano tensor of valence 2 and 
type $i$. We use for 
the Killing tensor $ K^{\mu\nu}_{ij}$ two additional indices $i,j$ 
to emphasize the fact that it is formed from two different Killing-Yano
tensors.

Introducing the Killing tensor $K^{\mu\nu}$ in the form (30) in the 
RHS of eq.(16) for
$n=1$ we can express the solution ${\cal J}^{(1)\mu}$ in terms of the
Killing-Yano tensors and their derivatives [15,32]:
\ba
{\cal J}^{(1)\mu}_{ij} &=& {i\over 2}\psi^\lambda \psi^\sigma (f^\nu_{i~\sigma}
D_\nu f^\mu_{j~\lambda} + f^\nu_{j~\sigma} D_\nu f^\mu_{i~\lambda}\no
&~&+ {1\over 2} f^{\mu\rho}_i c_{j\lambda\sigma\rho} +
 f^{\mu\rho}_j c_{i\lambda\sigma\rho} )
\ea
where the tensor $c_{i\mu\nu\lambda}$ is 
\be
c_{i\mu\nu\lambda} = -2 f_{i[\nu\lambda;\mu]}.
\ee

Finally, using ${\cal J}^{(1)\mu}_{ij}$ in the RHS of eq.(16) for $n=0$ 
we get for ${\cal J}^{(0)}_{ij}$
\be
{\cal J}^{(0)}_{ij} =-{1\over 4} \psi^\lambda\psi^\sigma\psi^\rho\psi^\tau
(R_{\mu\nu\lambda\sigma} f^\mu_{i~\rho} f^\nu_{j~\tau} + {1\over 2}
c^{~~~\pi}_{i\lambda\sigma} c_{j\rho\tau\pi}).
\ee

Collecting the quantities (30),(31) and (33) in eq.(24) we get the 
corresponding conserved quantity:
\be
{\cal J}_{ij} = {1\over 2!} K^{\mu\nu}_{ij}\Pi_\mu \Pi_\nu + 
{\cal J}^{(1)\mu}_{ij}\Pi_\mu + {\cal J}^{(0)}_{ij}.
\ee

Higher orders of the generalized Killing eq.(16) can be treated
similarly, but the corresponding expressions are quite involved. On the
other hand, for practical purposes (see Section 3), it turns out to be
sufficient to consider in detail the Killing tensors up to the valence 2. 

In what follows we shall return to the eq.(23)
looking for solutions depending on the Grassmann variables
{$\{$}$\psi^\mu${$\}$}. The existence of such kind of solutions of the
Killing equation is one of the specific features of the spinning particle 
models.

Even the lowest order eq.(23) with
$n=0$ has a non-trivial solution [17,31]
\be
{\cal J}^{(0)} = \frac i4 f_{\mu\nu}\psi^\mu\psi^\nu
\ee
where $f_{\mu\nu}$ is a Killing-Yano tensor
covariantly constant. Moreover, from eq.(24), we infer that 
${\cal J}^{(0)}$ is a separately conserved quantity.

The next eq.(23) with $n=1$ can have different kinds of solutions.
The most remarkable class of solutions is represented by
\be
{\cal J}_{\mu_1}^{(1)} = f_{\mu_1 \mu_2\dots\mu_r}\psi^{\mu_2}
\dots \psi^{\mu_r}
\ee
generated from a Killing-Yano tensor of valence $r$. 
Again, introducing this quantity in the RHS of eq.(16) for $n=0$
we get for ${\cal J}^{(0)}$:
\be
{\cal J}^{(0)} = \frac{i}{r+1}(-1)^{r+1} f_{[\mu_1\dots
\mu_r;\mu_{r+1}]} \cdot \psi^{\mu_1}\dots\psi^{\mu_{r+1}}
\ee
and the constant of motion corresponding to these solutions
of the generalized Killing equations is [31]:
\ba
Q_f &=& f_{\mu_1 \dots\mu_r}\Pi^{\mu_1}\psi^{\mu_2}\dots \psi^{\mu_r} \no
&~&+ \frac{i}{r+1}(-1)^{r+1}f_{[\mu_1 \dots
\mu_r;\mu_{r+1}]}
\cdot \psi^{\mu_1}\dots \psi^{\mu_{r+1}}.
\ea

This quantity is a superinvariant
\be
\{ Q_f , Q_ 0 \} = 0
\ee
for the bracket defined by eq.(13). 
A similar result was obtained in ref.[36] in which it is
discussed the role of the generalized Killing-Yano tensors, with
the framework extended to include electromagnetic interactions.
This result extends the analysis from refs.[15-17] where it is
established that the existence of a Killing-Yano tensor of the usual type
$(r=2)$ is both a necessary and a sufficient condition for the existence
of a new supersymmetry of the type (38) obeying the superinvariance
condition (39).

To conclude, we mention that eq.(23) with $n=1$ has also many other
solutions by forming combinations of different Killing-Yano tensors. As a
rule, the corresponding constants of motion are not completely new and
they can be expressed in terms of the quantities described above. We shall
illustrate this fact choosing for the solution of eq.(23) with $n=1$ the 
quantity
\be
{\cal J}^{(1)}_\mu = R_\mu f_{\lambda\sigma}\psi^\lambda
\psi^\sigma
\ee
where $R_\mu$ is a Killing vector (Killing-Yano tensor with $n=1$) and
$f_{\lambda\sigma}$ is a Killing-Yano tensor covariantly
constant. Introducing this solution in the RHS of the eq.(16)
with $n=0$, after some calculations, we get for ${\cal J}^{(0)}$
[31,32]:
\be
{\cal J}^{(0)} = \frac i2 R_{[\mu;\nu]}f_{\lambda\sigma}\psi^\mu 
\psi^\nu \psi^\lambda \psi^\sigma.
\ee

Combining eqs.(40) and (41) with the aid of eq.(24) 
we get the constant of motion :
\be
{\cal J} = f_{\mu\nu}\psi^\mu \psi^\nu \left( R_\lambda
\Pi^\lambda + \frac i2 R_{[\lambda;\sigma]}\psi^\lambda \psi^\sigma
\right).
\ee

As expected, we recognize in this expression the conserved quantities (27)
and (35).

In the next Section we shall apply these general results concerning the
solutions of the Killing equations for spinning spaces to the case of the
four-dimensional Euclidean Taub-NUT spinning space.
\newpage
\section{EUCLIDEAN TAUB-NUT SPINNING \\SPACE}
The Kaluza-Klein monopole [19,20] was obtained by embedding the Taub-NUT 
gravitational instanton into five-dimensional theory, adding the time 
coordinate in a trivial way. Its line element is expressed as:
\ba
ds^2_5&=&-dt^2+ds^2_4\no
&=&-dt^2+V^{-1}(r)[dr^2+r^2d\theta^2+r^2\sin^2\theta\,d\vf^2]\no
& &+V(r)[dx^5+\vec{A}(\vec{r})\,d\vec{r}\,]^2 
\ea
where $\vec{r}$ denotes a three-vector $\vec{r} =(r, \theta, \vf)$ and 
the gauge field $\vec{A}$ is that of a monopole 
\ba
A_r=A_\theta&=&0,~~~A_{\vf}=4m(1-\cos\theta)\no
\vec{B}&=&rot\vec{A}={4m\vec{r}\over r^3} .
\ea

The function V(r) is
\be
V(r)=\left(1+{4m\over r}\right)^{-1} 
\ee
and the so called NUT singularity is absent if $x^5$ is periodic with
period $16\pi m$ [37].

It is convenient to make the coordinate transformation
\be
4m(\chi+\vf)=-x^5
\ee
with $0\leq \chi < 4\pi$, which converts the four-dimensional line 
element $ds_4$ into
\be
ds^2_4=V^{-1}(r)[dr^2+r^2d\theta^2+r^2\sin^2\theta\, d\vf^2]
+16m^2 V(r)[d\chi+\cos\theta\, d\vf]^2 .
\ee

Spaces with a metric of the form given above have an isometry group
$ SU(2)\times U(1)$. The four Killing vectors are
\be
D_A=R_A^\mu\,\partial_\mu,~~~~A=0,1,2,3,
\ee
where
\ba
D_0&=&{\partial\over\partial\chi},\no
D_1&=&-\sin\vf\,{\partial\over \partial\theta}-\cos\vf\,\cot\theta
\,{\partial\over\partial\vf}+{\cos\vf\over\sin\theta}\,{\partial\over
\partial\chi},\no
D_2&=&\cos\vf\,{\partial\over \partial\theta}-\sin\vf\,\cot\theta
\,{\partial\over\partial\vf}+{\sin\vf\over\sin\theta}\,{\partial\over
\partial\chi},\no
D_3&=&{\partial\over\partial\vf}.
\ea

$D_0$ which generates the $U(1)$ of $\chi$ translations, commutes 
with the other Killing vectors. In turn the remaining three vectors,
corresponding to the invariance of the metric (47) under spatial
rotations ($A=1,2,3$), obey an $SU(2)$ algebra with
\be
[D_1, D_2]=-D_3~~~,{\it etc... }.
\ee

In the purely bosonic case these invariances would correspond to 
conservation of the so called ``relative electric charge" and the
angular momentum [18,23-25] :
\be
q=16m^2\,V(r)\,(\dot\chi+\cos\theta\,\dot\vf) ,
\ee
\be
\vec{j}=\vec{r}\times\vec{p}\,+\,q\,{\vec{r}\over r} .
\ee
where $\vec{p}=V^{-1}(r)\dot{\vec{r}}$ is the 
``mechanical momentum" which is only part of the momentum
canonically conjugate to $\vec{r}$.  

As observed in [18], the Taub-NUT geometry also possesses four Killing-Yano 
tensors of valence 2. The first three are rather special: they are
covariantly constant (with vanishing field strength)
\ba
f_i &=&8m(d\chi + \cos\theta d\varphi)\wedge dx_i -
\epsilon_{ijk}(1+\frac{4m}{r}) dx_j \wedge dx_k,\no
D_\mu f^\nu_{i\lambda} &=&0, ~~~~i=1,2,3.
\ea
Moreover, they are mutually anticommuting and square the minus unity:
\be
f_i f_j + f_j f_i = -2\delta_{ij}.
\ee

Thus they are complex structures realizing the quaternion algebra. Indeed,
the Taub-NUT manifold defined by (47) is hyper-K\" ahler and, as a
consequence, the corresponding supersymmetric $\sigma$-model has an $N=4$ 
supersymmetry.

The fourth Killing-Yano tensor is
\ba
f_Y &=&8m(d\chi + \cos\theta  d\varphi)\wedge dr \no &~&+
4r(r+2m)(1+\frac{r}{4m})\sin\theta  d\theta \wedge d\varphi
\ea
and has only one non-vanishing component of the field strength 
\be
{f_{Y}}_{r\theta;\varphi} = 2(1+\frac{r}{4m})r\sin\theta.
\ee

In the Taub-NUT case there is a conserved vector analogous to the
Runge-Lenz vector of the Kepler-type problem whose existence is rather
surprising in view of the complexity of the equations of motion. This
conserved vector is:
\be
\vec{K} = \frac 12 \vec{K}_{\mu\nu}\Pi^\mu\Pi^\nu = 
\vec{p}\times\vec{j} + \left(\frac{q^2}{4m}-
4mE\right)\frac{\vec r}{r}
\ee
where the conserved energy $E$, from eq. (10), is 
\ba
E&=&{1\over 2}\, g^{\mu\nu}\,\Pi_\mu\,\Pi_\nu\,
={1\over 2}V^{-1}(r)\left[\dot{\vec{r}}^{\,2} +\left( {q\over 
4m}\right)^2\right]  \no
&=&{1\over 2} {4m+r\over r}\,\dot r^2
+{1\over 2}(4m+r)\,r\,\dot\theta^2 +{1\over 2}(4m+r)\,r\,\sin^2\theta\,
\dot\vf^2\no
& &+ 8m^2\,{r\over 4m+r}\,(\cos\theta\,\dot\vf +\dot\chi)^2 .
\ea

The components $K_{i\mu\nu}$ involved with the Runge-Lenz type vector (57)
are Killing tensors and they can be expressed as symmetrized
products of the Killing-Yano tensors $f_i$ (53) and
$f_Y$ (55) as in eq. (30) [17,18,33]: 
\be
K_{i\mu\nu} = m\left( f_{Y\mu\lambda}
{{f_{i}}^\lambda}_\nu + f_{Y\nu\lambda} {{f_{i}}^\lambda}_\mu
\right) +\frac{1}{8m} (R_{0\mu} R_{i\nu} + R_{0\nu} R_{i\mu}).
\ee
This equation corrects some old formulas from the literature [18].

Using these conservation laws one can determine the orbits. Eq.(52)
implies that
\be
\vec{j}\cdot{\vec{r}\over r} = \vert \vec{j} \vert \cos\theta = q
\ee
which fixes the relative motion to lie on a cone whose vertex is at the
origin and whose axis is $\vec{j}$. Moreover, taking into account the
existence of the Runge-Lenz vector (57), one finds that the trajectories
lie simultaneously on the cone (60) and also in the plane perpendicular to
\be
\vec{n}=q\vec{{\cal K}}+\left( 4mE-{q^2\over 4m}\right)\vec{j}.
\ee
Thus they are conic sections.

Starting with these results from the bosonic sector of the Taub-NUT space
one can proceed with the spin contribution to the conserved quantities
(51),(52) and (57). 

First of all, corresponding to the {\it generic} symmetries described in
the previous Section, there are four universal conserved charges. For the
Taub-NUT spinning space these are :
\begin{enumerate}
\item{The energy (58);}
\item{The supercharge (19):
\ba
Q_0=& &{4m+r\over r}\dot r\,\psi^r+(4m+r)\,r\dot\theta\,\psi^\theta\no
&+&\left[ (4m+r)\,r\sin^2\theta\,\dot\vf+q\,\cos\theta\right]\,\psi^\vf 
+q\,\psi^\chi ; 
\ea}
\item{The chiral charge
\be\Gamma_* =4m(4m+r)r\,\sin\theta\,\psi^r\,\psi^\theta\,\psi^\vf
\,\psi^\chi ;
\ee}
\item{The dual supercharge
\ba
Q^*=& &4m(4m+r)r\,\sin\theta(\dot r\,\psi^\theta\,\psi^\vf\,\psi^\chi-
\dot\theta\,\psi^r\,\psi^\vf\,\psi^\chi\no
& &+\dot\vf\,\psi^r\,\psi^\vf
\,\psi^\chi  -\dot\chi\,\psi^r\,\psi^\theta\,\psi^\vf) .
\ea}
\end{enumerate}

From eq.(4) which shows that $\psi^\mu$ is covariantly constant, 
we find that the rate of change of the spins is:
\ba
\dot\psi^r&=&{2m\over r(4m+r)}\,\dot r\,\psi^r+{r^2+2mr\over 4m+r}
\,\dot\theta\,\psi^\theta\no
& &+\left({r^2+2mr\over 4m+r}\,\sin^2\theta
+{32m^3r\cos^2\theta\over (4m+r)^3}\right)\dot\vf\,\psi^\vf\no
& &+{32m^3r\cos\theta\over(4m+r)^3}(\dot\vf\,\psi^\chi+\dot\chi\,\psi^\vf)
+{32m^3r\over(4m+r)^3}\dot\chi\,\psi^\chi,\no
~\no
\dot\psi^\theta&=&-{r+2m\over r(4m+r)}(\dot r\,\psi^\theta+
\dot\theta\,\psi^r)+{8mr+r^2\over (4m+r)^2}\sin\theta\,\cos\theta\,\dot\vf
\,\psi^\vf\no
& &-{8m^2\sin\theta\over (4m+r)^2}(\dot\vf\psi^\chi+\dot\chi\psi^\vf),
\no
~\no
\dot\psi^\vf&=&-{r+2m\over r(4m+r)}(\dot r\,\psi^\vf+\dot\vf\,\psi^r)
-{8m^2+8mr+r^2\over (4m+r)^2}\,{\cos\theta\over\sin\theta}\,
(\dot\theta\,\psi^\vf\no
& &+\dot\vf\,\psi^\theta)
+ {8m^2\over (4m+r)^2}\,{1\over \sin\theta}\,(\dot\theta\,\psi^\chi+
\dot\chi\,\psi^\theta),\no
~\no
\dot\psi^\chi&=&{\cos\theta\over (4m+r)}(\dot r\,\psi^\vf+\dot\vf\,\psi^r)
-{2m\over r(4m+r)}(\dot r\,\psi^\chi+\dot\chi\,\psi^r)\no
& &+\left({8m^2+8mr+r^2\over (4m+r)^2}{\cos^2\theta\over\sin\theta}
+{1\over 2}\sin\theta\right)(\dot\theta\,\psi^\vf+\dot\vf\,\psi^\theta)
\no
& &-{8m^2\over (4m+r)^2}\,{\cos\theta\over\sin\theta}\,(\dot\theta\,
\psi^\chi+\dot\chi\,\psi^\theta).
\ea

As a rule, the complicated eqs.(3) and (4) should be integrated to 
obtain the full solution of the equations of motion for the usual 
coordinates $\{$$ x^\mu$$\}$ and Grassmann coordinates 
$\{$$ \psi^\mu$$\}$. In addition to the brute force method of trying to
solve these equations there are the general prescriptions described in the
previous Section which are considerable simpler. In what follows we shall
use the Killing-Yano tensors to generate the constants of motion 
for spinning particles.

We start with the observation that the angular momentum (52) and 
the ``relative electric charge'' (51) are
constructed with the aid of the Killing vectors (49). The corresponding
conserved quantities in the spinning case are the followings:
\be
{\vec J}={\vec B} + {\vec j},
\ee
\be
J_0 = B_0 + q
\ee
where we used eq.(27), and we introduced the notation: 
${\vec J} =(J_1, J_2, J_3), {\vec B} =
(B_1, B_2, B_3)$. From eq.(26), the scalars $B_A$ 
\be
B_{A} = \frac i2 R_{A[\mu;\nu]}\psi^\mu \psi^\nu
\ee
have the following  detailed expressions: 

\ba
B_0=& &\frac{32 m^3 \cos{\theta}}{(4m+r)^2} S^{r\vf} +\frac{32
m^3}{(4m+r)^2} S^{r\chi} -\frac{8m^2 r\sin{\theta}}{4m+r} S^{\theta\vf},\no
&~&\no
B_1=& &-\sin\vf \left( (2m+r)S^{r\theta} +\frac{8m^2 r\sin{\theta}}{4m+r} 
S^{\vf\chi} \right) \no
&+&\cos\vf\left[ \left(\frac{32m^3}{(4m+r)^2}-(2m+r)\right)
\sin{\theta}\cos\theta S^{r\vf}\right.\no
&+ &\left.\frac{32m^3 \sin\theta}
{(4m+r)^2} S^{r\chi}+ \frac{8m^2r+(8mr^2 +r^3)\sin{\theta}^2}{4m+r}
\right.\no
&+ &\left.\frac{8m^2r\cos{\theta}}{4m+r}S^{\theta\chi}\right],\no
&~&\no
B_2=& &\cos\vf\left((2m+r)S^{r\theta} +\frac{8m^2 r\sin{\theta}}{4m+r} 
S^{\vf\chi}\right)\no
&+&\sin\vf\left[ \left(\frac{32m^3}{(4m+r)^2}-(2m+r)\right)
\sin{\theta}\cos\theta S^{r\vf}\right.\no
&+ &\left.\frac{32m^3 \sin\theta}
{(4m+r)^2} S^{r\chi}+\frac{8m^2r+(8mr^2 +r^3)\sin{\theta}^2}{4m+r}
\right.\no
&+ &\left.\frac{8m^2r\cos{\theta}}{4m+r}S^{\theta\chi}\right],\no
&~&\no
B_3=& &\left[(2m+r) \sin{\theta}^2 +\frac{32m^3\cos{\theta}^2}
{(4m+r)^2}\right]S^{r\vf} +\frac{32m^3 \cos\theta}{(4m+r)^2}S^{r\chi}\no
&+&\frac{(8mr^2 +r^3)\sin\theta\cos\theta}{4m+r}S^{\theta\chi}
-\frac{8m^2 r}{4m+r}\sin\theta S^{\theta\chi}.
\ea

We mention that the above constants of motion are superinvariant:
\be
\left\{ J_A , Q_0 \right\} = 0~~~,~~~A=0,\dots,3.
\ee
Also, the components (66) of the angular momentum satisfy, as
expected, the $SO(3)$ algebra:
\be
\left\{ J_{i} , J_{j} \right\} = \epsilon_{ijk} J_k~~~,~~~
i,j,k=1,2,3.
\ee

We consider now the Killing-Yano tensors of valence 2 and we search for 
those constants of motion built of them.

Using eq.(38) we can construct from the Killing-Yano tensors (53) and (55)
the supercharges $Q_i$ and $Q_Y$. The supercharges $Q_i$ together $Q_0$
from eq.(19) realize the $N=4$ supersymmetry algebra [17]:
\be
\left\{ Q_A , Q_B \right\} = -2i\delta_{AB}H~~~,~~~A,B=0,\dots,3
\ee
making manifest the link between the existence of the
Killing-Yano tensors (53) and the hyper- K\" ahler geometry of the
Taub-NUT manifold. Moreover, the supercharges $Q_i$ transform as vectors 
at spatial rotations
\be
\{Q_i ,J_j\}=\epsilon_{ijk} Q_k~~~,~~~i,j,k=1,2,3
\ee
while $Q_Y$ and $Q_0$ behave as scalars.

We note also that the bracket of $Q_Y$ with itself can be 
expressed in terms of the hamiltonian, angular momentum 
and ``relative electric charge'':
\be
\{Q_Y , Q_Y\}=-2i\left( H +\frac{{\vec J}^{~2} 
- {J_0}^2}{4 m^2}\right).
\ee

On the other hand, the existence 
of the Killing-Yano covariantly constant tensors $f_i$ (53) is connected  
with three new constants of motion as shown in eq.(35):
\be
S_{i} = \frac i4 f_{i\mu\nu}\psi^\mu \psi^\nu
~~~,~~~ i=1,2,3
\ee
which realize an $SO(3)$ Lie-algebra similar to that of the
angular momentum (71):
\be
\left\{S_{i} , S_{j}\right\} = \epsilon_{ijk}
S_{k}~~~,~~~i,j,k = 1,2,3.
\ee

These components of the spin are separately conserved and can be
combined with the angular momentum $\vec{J}$ to define a new
improved form of the
angular momentum $I_{i} = J_{i} - S_{i}$ with the property that it 
preserves the algebra [17]:
\be
\left\{ I_{i},I_{j} \right\} = 
\epsilon_{ijk}I_{k}~~~,~~~i,j,k = 1,2,3
\ee
and that it commutes with the $SO(3)$ algebra generated by the
spin $S_{i}$ 
\be
\left\lbrace I_{i},S_{j} \right\rbrace = 0.
\ee

Let us note also the following Dirac brackets of $S_{i}$ with 
supercharges
\ba
\left\{ S_{i},Q_0 \right\} &=& -\frac{Q_{i}}{2},\no
\left\{ S_{i},Q_{j} \right\}&=& \frac1 2 (\delta_{ij} Q_0
+ \epsilon_{ijk}Q_{k}).
\ea

To get the spin correction to the Runge-Lenz vector (57) it is
necessary to investigate the generalized Killing eqs.(16)
for $n=1$ with the Killing tensor ${\vec K}_{\mu\nu}$  in
the RHS. For an analytic expression of the solution of this equation 
we shall use the decomposition (59) of the Killing tensor 
${\vec K}_{\mu\nu}$ in terms of Killing-Yano tensors.
Starting with this decomposition of the Runge-Lenz vector 
${\vec K}$ from the scalar case, it is possible to express the
corresponding conserved quantity ${\vec {\cal K}}$ in the 
spinning case [33]:
\be
{\cal K}_i = 2m \left( -i\{ Q_Y , Q_i\} + \frac{1}{8 m^2} J_i
J_0 \right)
\ee

This expression differs from previous results presented in the
literature [17,31] and the difference has the origin in the
corrected form of relation (59). A detailed expression of the
components ${\cal K}_{i\mu\nu}$ is:
\ba
{\cal K}_i &=& 2m \left[ \left((f_Y f_i)_{(\mu\nu)} +\frac{1}
{8 m^2} R_{i(\mu} R_{0\nu)}\right) \Pi^\mu \Pi^\nu \nonumber
\right.\\
& &+\left.\biggl( {{f_i}^\lambda}_\beta f_{Y\mu\alpha;\lambda} +
{{f_i}^\lambda}_\mu f_{Y\alpha\beta;\lambda} \right.\biggr.
\no
& &- \left.\left.\frac{1}{16 m^2}(R_{i\alpha;\beta} R_{0\mu} 
+ R_{0\alpha;\beta} R_{i\mu})\right)S^{\alpha\beta}\Pi^\mu \right.
\no
& &+\left.\frac{1}{32 m^2} S^{\alpha\beta} S^{\gamma\delta}
R_{i\alpha;\beta} R_{0\gamma;\delta}\right].
\ea

More explicitly, we can write the Runge-Lenz vector ${\vec{\cal K}}$
for the spinning case as in eq.(34):
\be
\vec{{\cal K}} =\frac 12 \vec{K}_{\mu\nu}\Pi^\mu \Pi^\nu 
+ \vec{S}_\mu \Pi^\mu + {\cal \vec S}
\ee
where the first term is the Runge-Lenz vector (57) from the scalar case
and the last two terms represent the specific spin contribution. 
A detailed expression of this contribution is [30,33]:
\ba
{S_1}_\mu \cdot \Pi^\mu&=&\left[-(4m+r)\cos\theta\cos\vf\cdot 
S^{r\theta}\right. \no
&&+\left. 4mr\cos\theta\sin\vf\cdot S^{\theta\vf}
+4mr\sin\vf\cdot S^{\theta\chi}\right.\no
&&+\left. (4m+r)\sin\theta\sin\vf\cdot S^{r\vf}+
4mr\sin\theta\cos\theta\cos\vf \cdot S^{\vf\chi}\right] \dot{r}\no
&&+\left[r(4m+r)\sin\theta\cos\vf\cdot S^{r\theta}
-{4mr(6m+r)\over 4m+r}\cos\theta\sin\vf\cdot S^{r\vf}\right.\no
&&-\left.{4mr(6m+r)\over 4m+r}sin\vf\cdot S^{r\chi} \right.\no
&&+\Biggl.r^2(6m+r)\sin\theta\sin\vf S^{\theta\vf}-
4mr^2\sin^2\theta\cos\vf S^{\vf\chi} \Biggr] \dot{\theta}\no
&&+\left[{4mr(6m+r)\over 4m+r}\cos\theta\sin\vf S^{r\theta}\right.\no
&&+\left(r(4m+r)\sin^3\theta\cos\vf+
{256m^4r\over(4m+r)^3}\sin\theta\cos^2\theta\cos\vf\right) S^{r\vf}\no
&&-\left(4mr+{8m^2r\over 4m+r}-{256m^4r\over(4m+r)^3}\right)\sin\theta
\cos\theta\cos\vf S^{r\chi}\no
&&+\left({r^2(32m^3+64m^2r+14mr^2+r^3)\over(4m+r)^2}\sin^2\theta\cos
\theta\cos\vf\right.\no 
&&+\left. {32m^3r^2\over(4m+r)^2}\cos^3\theta\cos\vf\right) S^{\theta
\vf}\no
&&+\left(4mr^2\sin^2\theta\cos\vf+{32m^3r^2\over (4m+r)^2}\cos^2\theta
\cos\vf\right) S^{\theta\chi}\no
&&-\left.{32m^3r^2\over (4m+r)^2}\sin\theta\cos\theta\sin\vf S^{\vf\chi}
\right]\dot{\vf}\no
&&+\left[ {4mr(6m+r)\over 4m+r}\sin\vf S^{r\theta}+{256m^4r\over
(4m+r)^3}\sin\theta\cos\vf S^{r\chi}\right.\no
&&+\left(4mr +{8m^2r\over 4m+r} +{256m^4r\over (4m+r)^3}\right)\sin\theta
\cos\theta\cos\vf S^{r\vf}\no
&&+\left({32m^3r^2\over (4m+r)^2}\cos^2\theta\cos\vf \right.\no
&&-\left.2 {64m^3r^2+16m^2r^3+2mr^4\over (4m+r)^2}\sin^2\theta\cos\vf\right) 
S^{\theta\vf}\no
&&+\left.{32m^3r^2\over (4m+r)^2}\cos\theta\cos\vf S^{\theta\chi}-
{32m^3r^2\over(4m+r)^2}\sin\theta\sin\vf S^{\vf\chi}\right]\dot{\chi}\no
&~&\no
{\cal S}_1 &=& \frac{8m^2r^2(8m+r)\sin{\theta}^2\cos\vf}{(4m+r)^3}S^{r\theta}
S^{\vf\chi} ,
\ea

\ba
{S_3}_\mu \cdot \Pi^\mu&=&\left[(4m+r)\sin\theta S^{r\theta}
-4mr\sin^2\theta S^{\vf\chi}\right]\dot{r}\no
&&+\left[r(4m+r)\cos\theta S^{r\theta}-4mr^2\sin\theta\cos\theta
S^{\vf\chi}\right]\dot{\theta}\no
&&+\left[\left({256m^4r\over (4m+r)^3}\cos^3\theta+r(4m+r)
\sin^2\theta\cos\theta\right)S^{r\vf}\right.\no
&&+\left({256m^4r\over (4m+r)^3}\cos^2\theta+{4mr(6m+r)\over 4m+r}
\sin^2\theta\right) S^{r\chi}\no
&&-\left(r^2(6m+r)\sin^3\theta +{96m^3r^2\over (4m+r)^2}
\sin\theta\cos^2\theta\right) S^{\theta\vf}\no
&&+\left. 2mr^2\left(2-{16m^2\over (4m+r)^2}\right)\sin\theta\cos\theta
S^{\theta\chi}\right]\dot{\vf}\no
&&+\left[\left(-4mr\sin^2\theta-{8m^2r\over 4m+r}\sin^2\theta
+{256m^4r\over (4m+r)^3}\cos^2\theta\right) S^{r\vf}\right.\no
&&+{256m^4r\over (4m+r)^3}\cos\theta S^{r\chi}-
{32m^3r^2\over (4m+r)^2}\sin\theta S^{\theta\chi}\no
&&-\left.4mr^2\left(1 + {24m^2\over (4m+r)^2}\right)\sin\theta\cos\theta
S^{\theta\vf}\right]\dot{\chi}\no
&~&\no
{\cal S}_3 &=&\frac{8m^2r^2(8m+r)\cos\theta\sin\theta}{(4m+r)^3}
S^{r\theta}S^{\vf\chi}.
\ea

The components ${S_2}_\mu \Pi^\mu$ and ${\cal S}_2$ can be obtained from
${S_1}_\mu \Pi^\mu$ respectively ${\cal S}_1$ with the substitutions:
\ba
\sin\vf\longrightarrow &-&\cos\vf,\no
\cos\vf\longrightarrow & &\sin\vf.
\ea

Therefore, in the spinning case, the Runge-Lenz
vector contains additional terms linear and quadratic in the
spin. The presence of a contribution quadratic in the spin,
non-existent in Refs.[17,30], is again related to the term
$J_i J_0$ from eq.(80). We would like to emphasize that the
contribution of the $J_i J_0$ term in eq.(80) is essential in
reproducing the known vectorial expression of the Runge-Lenz
vector in the scalar case, and gives the correct Poisson-Dirac
bracket between two components of the Runge-Lenz vector.

The Dirac brackets involving the Runge-Lenz vector (80) are
(after some algebra):
\ba
\{ {\cal K}_i , Q_0 \} &=& 0,\no
\{ {\cal K}_i , J_j\} &=& \epsilon_{ijk} {\cal K}_k,\no
\{ {\cal K}_i , {\cal K}_j \} &=& \epsilon_{ijk} J_k \left[
\frac{{J_0}^2}{16 m^2} -2H\right]
\ea
and they are similar to those known from the scalar case.
The Runge-Lenz vector $\vec{\cal K}$
together with the total angular momentum $\vec{J}$
generates an $SO(4)$ or $SO(3,1)$ algebra depending upon the sign 
of the quantity $\left(\frac{{J_0}^2}{16 m^2}-2E\right)
\vert~_{\psi^\mu=0}$ is positive or negative.

In conclusion, all conserved quantities for motions in spinning Taub-NUT
space have been expressed in terms of the geometric objects (49), (53) and
(55). Other expressions involving the Killing-Yano tensors will produce
conserved quantities which are not new, 
but rather combinations of the above primary conserved quantities. 
For example, let us consider a solution of the homogeneous eq.(23)
for $n=1$ of the type (40):
\ba
{\cal J}^{(1)}_{Aj\mu} &=&
R_{A\mu} f_{j \lambda\sigma}\psi^\lambda \psi^\sigma, \no
&~&A=0,\dots,3~~~,~~~j=1,2,3.
\ea

After some algebra we get the constants of motion of the
form (42):
\ba
{\cal J}_{Aj} &=& f_{j_ \lambda\sigma}\psi^\lambda \psi^\sigma
\left( R_{A \mu}\Pi^\mu + \frac i2  R_{A [\alpha;\beta]}
\psi^\alpha \psi^\beta\right) \no
&=& -4i S_{j} J_{A} ~~~,~~~ A=0,\dots,3~~~,~~~j=1,2,3.
\ea

As expected, the constants ${\cal J}_{Aj}$ are not 
new, being expressed in terms of the constants 
$J_{A}$ (62), (63) and $S_{j}$ (75). However, the combinations (88) 
arise in a natural way as solutions of the generalized Killing 
equations and appear only in the spinning case. Moreover, we can 
form a sort of Runge-Lenz vector involving only Grassmann 
components:
\be
L_{i} = \frac 1m \epsilon_{ijk} S_{j} J_{k}~~~,~~~i,j,k=1,2,3
\ee
with the commutation relations like in eqs.(86):
\ba
\{ L_i , J_j \} &=& \epsilon_{ijk} L_{k},\no
\{ L_i , L_j \} &=& \left( \vec{S}\vec{J} - \vec{S}^2
\right)\frac{1}{m^2} \epsilon_{ijk} J_{k}.
\ea

Note also the following Dirac brackets of $L_i$ with supercharges:
\ba
\{ L_i , Q_0 \} &=& -\frac{1}{2m} \epsilon_{ijk}
Q_{j} J_{k},\\
\{ L_{i} , Q_{j} \} &=& \frac{1}{2m} (\epsilon_{ijk} Q_0 
J_k - \delta_{ij} Q_{k} M_{k}^{-} 
+ Q_{i} M_{j}^{-}).
\ea
\newpage
\section{SPECIAL SOLUTION}
In spite of the fact that all conserved quantities have been expressed in
terms of geometric ones in a close form, their detailed expressions (69),
(83), (84) are quite intricate.

We wish to consider a special class of solution of the equations of motion
which is very simple, but not at all trivial. In this Section we confine
ourselves to the motion on a cone on the analogy of the scalar case where
the trajectories are conic sections.

For this purpose let us choose the $z$ axis along $\vec{J}$ so that the 
motion of the particle may be conveniently described in terms of
polar coordinates
\be
\vec{r}=r\vec{e}\,(\theta,\vf)
\ee
with
\be
\vec{e}=(\sin\theta\cos\vf, \sin\theta\sin\vf, \cos\theta).
\ee

For this choice of the axis we have:
\be
(4m+r)r\dot\theta=-(2m+r) S^{r\theta}-\frac{8m^2
r}{4m+r}\sin{\theta} S^{\vf\chi},
\ee

\ba
\dot{\vf}=& &\frac{q}{r(4m+r)\cos{\theta}}-
\left[ \left(
\frac{32m^3}{r(4m+r)^3}-\frac{2m+r}{r(4m+r)}\right) S^{r\vf}\right.\no
&+&\left.\frac{32m^3}{r(4m+r)^3\cos\theta} S^{r\chi}
+\frac{8m^2+(8mr+r^2)\sin{\theta}^2}{(4m+r)^2 \sin\theta
\cos\theta}S^{\theta\vf} \right.\no
&+&\left.\frac{8m^2}{(4m+r)^2\sin\theta} S^{\theta\chi}\right]
\ea
and from eqs.(66), (67) and (69)
\be
J_0-\frac{\vec J\vec r}{r}=-r(4 m+r)S^{\theta\vf}\sin\theta.
\ee

In what follows we shall consider the angle $\theta = constant$, i.e. 
$\dot\theta=0$. The rate of change of spin (65) can be written in terms of
the antisymmetric spin polarization tensor $S^{\mu\nu}$ (5) as follows:
\ba
\dot{S}^{r\theta}=&-&{\dot r\over 4m+r}\,S^{r\theta}+{8m+r\over 
2(4m+r)^3}\,q\,\sin\theta \,S^{r\vf}\no
&-&{8m^2\over r(4m+r)^3}\,{q\sin\theta
\over\cos\theta} \,S^{r\chi}
-{r\sin^2\theta+2m\over(4m+r)^2}\,{q\over\cos\theta}\,S^{\theta\vf}\no
&-&{2mq\over (4m+r)^2} \,S^{\theta\chi},\no
~\no
\dot{S}^{r\vf}=&-&{q\over2r(4m+r)\sin\theta} \,S^{r\theta}-{\dot r
\over 4m+r}\,S^{r\vf} -{2mq\over (4m+r)^2} \,S^{\vf\chi},\no
~\no
\dot{S}^{r\chi}=& &{q\over 2r(4m+r)\sin\theta\cos\theta}\,S^{r\theta}
+{\dot {r}\cos\theta\over 4m+r} \,S^{r\vf}\no
&+&{(r\sin^2\theta+2m)\,q\over
(4m+r)^2\cos\theta} \,S^{\vf\chi},\no
~\no
\dot{S}^{\theta\vf}&=&{(2m+r)\,q\over r^2 (4m+r)^2\cos\theta} \,S^{r\theta}
-2{2m+r\over r(4m+r)}\,\dot {r}\,S^{\theta\vf}\no
&+&{8m^2\sin\theta \,q\over r
(4m+r)^3\cos\theta} \,S^{\vf\chi},\no
~\no
\dot{S}^{\theta\chi}&=&{q\over 8m(4m+r)^2} \,S^{r\theta}+{\dot {r}
\over 4m+r}\cos\theta \,S^{\theta\vf}-{\dot r\over r} \,S^{\theta\chi}
\no
&+&{8m+r\over 2(4m+r)^3}\,q\sin\theta \,S^{\vf\chi},\no
~\no
\dot{S}^{\vf\chi}&=&{q\over 8m(4m+r)^2} \,S^{r\vf}-{(2m+r)\,q\over
r^2(4m+r)^2\cos\theta}\,S^{r\chi}\no
&-&{q\over 2r(4m+r)\sin\theta\cos\theta} \,S^{\theta\vf}-{q\over
2r(4m+r)\sin\theta} \,S^{\theta\chi}\no
&-&{\dot r \over r} \,S^{\vf\chi}.
\ea

Since we are looking for solutions with $\dot\theta =0$ we have from
eq.(95)
\be
S^{r\theta}+\frac{8 m^2 r \sin\theta}{(2m+r)(4m+r)}S^{\vf\chi}=0.
\ee

This relation implies that the special solution investigated in this
Section is situated in the sector with
\be 
\Gamma_*=0.
\ee

We mention that in this sector the system of eqs.(98) is satisfied even if
the angle $\theta$ is not constant.

Using eq.(99) we can express $S^{r\theta}$ through $S^{\vf\chi}$ and the
following equations are equivalent to the system (98)
\ba
{d\over dt}\left[ (4m+r) S^{r\vf}\right]&=&{r\over 4m+r}\,q S^{\vf\chi},
\no
~\no
{d\over dt}\left[ \cos\theta S^{r\vf}+ S^{r\chi}\right]&=&0,
\no
~\no
{d\over dt}\left[ r(4m+r) S^{\theta\vf}\right]&=&-2{\sin\theta\over
\cos\theta}\,{r\over 4m+r}\,q \,S^{\vf\chi},
\no
~\no
{d\over dt}\left[ r\cos\theta \,S^{\theta\vf}+ r\,S^{\theta\chi}\right]&=&
-{\sin\theta\over 4m}\,{r\over 4m+r}\,q \,S^{\vf\chi}.
\ea

Thus the equations of motion for $S^{\mu\nu}$ are written in a more
tractable form and the solution follows without difficulties. If we take into
consideration the constraint coming from eq.(97), namely:
\be
\frac{d}{dt}\left[r(4m+r)S^{\theta\vf}\right]=0
\ee
then we have to impose 
\be
q\cdot S^{\vf\chi} = 0.
\ee
on the system (101). We shall analyze both solutions 
$S^{\vf\chi} = 0$ and $q = 0$ successively.

For  $S^{\vf\chi} = 0$, from eq.(99) we have also
\be
S^{r\theta}  = 0.
\ee

In spite of this drastic simplification, eqs.(101) have a non-trivial
solution:
\ba
S^{r\vf}&=&\frac{(\sin\theta \mp 1)}{\cos\theta(4m+r)}
\Sigma ,\no
S^{r\chi}&=&\frac{\sin\theta}{4m}\Sigma
-\frac{(\sin\theta \mp 1)}{4m+r}\Sigma,\no
S^{\theta\vf}&=&\frac{1}{r(4m+r)}\Sigma,\no
S^{\theta\chi}&=&\frac{-(\sin\theta\pm
1)\tan\theta}{4mr} \Sigma -\frac{\cos\theta}{r(4m+r)} \Sigma
\ea
where $\Sigma$ is a Grassmann constant, commuting with ${\psi^\mu}$, and 
anticommuting with itself.

In the case of this particular solution, from eqs.(67)-(69) we get that
the spin contribution to the ``relative electric charge'' vanishes 
($B_0 = 0$) and
\be
J_0 = q.
\ee
Therefore the ``relative electric charge'' has the same expression as in
the scalar case. However the total angular momentum is modified by the
spin contribution:
\be
J_0-\frac{\vec J\vec r}{r}= q - J\cos\theta = -\Sigma\sin\theta.
\ee
Here $J$ is the magnitude of the total angular momentum and eq.(107) fixes
the angle $\theta$ in terms of the constants $q, J $ and $\Sigma$.
Also the equations for $\vf$ and $\chi$ are modified:
\ba
\dot\vf&=&\frac{q}{r(4m+r)\cos\theta} \pm
\frac{\Sigma}{(4m+r)^2\cos\theta},\no
\dot\chi&=&\frac{8m+r}{16m^2(4m+r)}q \mp
\frac{\Sigma}{(4m+r)^2}.
\ea
At last, $\dot r$ can be derived from the energy, eq.(58).

Concerning the second possibility, namely $q=0$ in eq.(103), 
we have  from eqs.(101):
\ba
S^{r\vf}&=&{{\cal C}^{r\vf}\over 4m+r},\no
S^{r\chi}&=&{\cal C}^{r\chi}\,-\,{\cos\theta\over 4m+r}\,
{\cal C}^{r\vf},\no
S^{\theta\vf}&=&{{\cal C}^{\theta\vf}\over r(4m+r)},\no
S^{\theta\chi}&=&{{\cal C}^{\theta\chi}\over r}\,-\,\cos\theta\,
{{\cal C}^{\theta\vf}\over r(4m+r)},\no
S^{r\theta}&=&\frac{C^{r\theta}}{4m+r},\no
S^{\vf\chi}&=&\frac{C^{\vf\chi}}{r},
\ea
where $C^{\mu\nu}$ are Grassmann constants of the same kind as $\Sigma$.

Taking into account the constraints:
\be
J_0=B_0=\frac{8m^2(4mC^{r\chi}-\sin\theta C^{\theta\vf})}{(4m+r)^2},
\ee
and, from (99), where we substitute (109):
\be 
C^{r\theta}+\frac{8m^2\sin\theta C^{\vf\chi}}{2m+r}=0
\ee
we get
\be
C^{r\theta}=0~~~,~~~C^{\vf\chi}=0
\ee
and
\be
C^{r\chi}=0~~~,~~~C^{\theta\vf}=0
\ee
or
\be
C^{r\chi}=\frac{\sin\theta}{4m}C^{\theta\vf}.
\ee
In conclusion, the case $q=0$ is included into the previous case
($S^{\vf\chi}$ = 0). Practically we must impose in addition to the
previous solution the condition that $q = 0 $ and we get either the solution 
(112), (114) or a trivial one without any spin contribution (112), (113).

Concerning the Runge-Lenz vector for $\theta = constant $ we have from
eqs.(83) and (84):
\ba
S_{1\mu}\Pi^\mu&=& \mp 2\tan\theta\sin\vf\dot r\Sigma
\mp 2r\tan\theta\sin^2{\theta}\cos\vf\dot\vf \Sigma \no
&&\mp\frac{(8m+r)\sin\theta\cos\vf}{4m(4m+r)}q\Sigma,
~\no
S_{3\mu}\Pi^\mu &=& \mp 2r\sin^2{\theta}\dot\vf\Sigma\no
&&\pm\frac{\tan\theta((8m+r)\sin\theta
\mp(4m+r))}{4m(4m+r)}q\Sigma .
\ea
Again the component $S_{2\mu}\Pi^\mu$ can be obtained from 
$S_{1\mu}\Pi^\mu$ with the substitution (85).

Using then:
\be
\vec S\vec p=-\frac{(4m+r)\dot r\sin\theta q\Sigma}{4mr},
\ee
\be 
\frac{\vec p\vec r}{r}=\frac{(4m+r)\dot r}{r},
\ee
\be
\vec J\vec p=(q+\Sigma\sin\theta)\frac{\vec p\vec r}{r}.
\ee
we get to the conclusion that, for the case of motion lying on a cone,
there is a conserved vector $\vec n$ orthogonal on
$\vec p$:
\be
\vec n=q\vec{\cal K}+\frac{(q-\Sigma\sin\theta)\vec J}{q}\left(4mE
-\frac{q(q-\Sigma\sin\theta)}{4m} \right)
\ee
which is similar to the scalar case, eq.(61).
Physical observables are obtained by averaging with some suitable density
over the anticommuting parameters. After the integration was performed 
we may treat $\Sigma$ as a classical variable. Therefore the trajectories
of a spinning particle constrained to the motion on a cone are conic sections
determined by the condition $\vec n\vec p =0$.
\newpage
\section{CONCLUDING REMARKS}
The spinning particle model is a world line supersymmetric
extension of the theory of a scalar particle. It describes a relativistic
particle with spin one half.

It is a theory which describes in a pseudo-classical way a Dirac particle
moving in an arbitrary $d$-dimensional space-time. In addition to the
usual space-time coordinates, the model involves anticommuting vectorial
coordinates which take into account the spin degrees of freedom. It is
worth emphasizing that along the world line of the particle there is a
supersymmetry between the fermionic spin variables and the bosonic
position coordinates. The model is a one-dimensional supersymmetric 
field theory on the world line. 

The term pseudo-classical refers to the fact that there is no classical
interpretation for the anticommuting variables. To get the
"observable" trajectories one has to average over the spin variables.
On the other hand it is possible to quantize the model giving rise to 
supersymmetric quantum mechanics. After quantization the conservation law
for the supercharge becomes the Dirac equation.

The constants of motion of a scalar particle in a curved space-time are 
determined by the symmetries of the manifold, i.e. if a space-time admits
a Killing tensor $K_{\mu_1\cdots\mu_r}$ of valence $r$, then the quantity
$K_{\mu_1\cdots\mu_r}\Pi^\mu_1\cdots\Pi^\mu_r$ is conserved along the 
geodesic.

In the spinning case the generalized Killing equations (16) are more
involved and new procedures should be conceived. The aim of this paper was 
to point out the important role of the Killing-Yano tensors to generate 
solutions of the generalized Killing equations. We presented a detailed 
discussion on how to construct conserved quantities out of Killing-Yano 
tensors for the Taub-NUT spinning space. Finally we solved the equation of 
motion for the case when the angle $\theta$ is held fixed. This solution
is most simple but far from trivial and the trajectories are the analogous of
the ones of a scalar particle, being conic sections. 

The extension of these results for the motion of spinning particles 
in spaces with torsion [38] and/or interacting with background
fields [3,36] will be discussed elsewhere [39].
\subsection*{Acknowledgements}~~
One of the authors (M.V.) would like to thank M.S.Marinov and M.Moshe for
useful discussions on the spinning spaces.
%
%

%
\end{document}